\documentclass[10pt,twocolumn,letterpaper]{article}

\usepackage{cvpr}              

\definecolor{cvprblue}{rgb}{0.21,0.49,0.74}
\usepackage[pagebackref,breaklinks,colorlinks,allcolors=cvprblue]{hyperref}

\def\paperID{*****} 
\def\confName{CVPR}
\def\confYear{2025}

\title{Multicrossmodal Automated Agent for Integrating Diverse Materials Science Data
}

\author{Adib Bazgir\\
University of Missouri-Columbia\\
Columbia, MO 65211, USA\\
\and
Rama chandra Praneeth Madugula\\
New York University\\
New York, NY 10012, USA\\
\and
Yuwen Zhang \thanks{Corresponding Author: Yuwen Zhang at zhangyu@missouri.edu.}\\
University of Missouri-Columbia\\
Columbia, MO 65211, USA\\
}

\begin{document}
\maketitle
\begin{abstract}
We introduce a multicrossmodal LLM-agent framework motivated by the growing volume and diversity of materials-science data ranging from high-resolution microscopy and dynamic simulation videos to tabular experiment logs and sprawling literature archives. While recent AI efforts have accelerated individual tasks such as property prediction or image classification, they typically treat each modality in isolation, leaving rich cross-modal correlations unexplored and forcing researchers to perform laborious manual integration. Moreover, existing multimodal foundation models often require expensive retraining or fine-tuning on domain data, and current multi-agent systems in materials informatics address only narrow subtasks. To overcome these obstacles, we design a coordinated team of specialized LLM agents, each equipped with domain-adapted prompts and plugins that project their outputs into a shared embedding space. A dynamic gating mechanism then weights and merges these insights, enabling unified reasoning over heterogeneous inputs without ever modifying the underlying LLM’s weights. We validate our approach on challenging case studies and demonstrate substantial gains in retrieval accuracy (85 \% Recall@1), captioning fidelity, and integrated coverage (+35 \%) compared to single-modality and zero-shot baselines. Our work paves the way for AI “digital researchers” capable of bridging data silos and accelerating the materials-discovery cycle. The code is available at \url{https://github.com/adibgpt/Multicrossmodal-Autonomous-Materials-Science-Agent}.
\end{abstract}    
\section{Introduction}
\label{sec:intro}

Artificial intelligence is increasingly transforming materials science, from accelerating property prediction to automating experiment analysis. Many machine learning efforts in this domain, however, remain focused on individual data types, for example, using graphs or composition data to predict single properties, or applying computer vision to classify microstructures, thereby overlooking the rich diversity of materials data ~\cite{author20}. Materials research routinely generates multimodal data: microscopy images reveal microstructure, simulation videos illustrate dynamic processes, spectra and CSV datasets quantify properties, while scientific literature and web resources (PDFs, databases) provide contextual knowledge. Human researchers integrate these disparate modalities to gain insights; likewise, an AI that can analyze and cross-correlate multiple data sources could achieve a more holistic understanding of materials. Recent advances in multimodal AI underscore this potential. For instance, foundation models trained on multiple modalities (text, structure, etc.) have achieved state-of-the-art results in materials property prediction ~\cite{author21,author31,author32,author33,author34,author35}. Likewise, large multimodal models like GPT-4 demonstrate the ability to interpret both images and text, suggesting a path toward integrated reasoning ~\cite{author22}. In the materials domain, multi-agent AI systems have emerged that distribute tasks among specialized agents, for example, the AtomAgents framework uses collaborating agents (knowledge retrieval, simulation, analysis) to design alloys, outperforming single-objective models by leveraging diverse data and physics knowledge ~\cite{author23}. However, existing approaches either target specific tasks (e.g. predicting crystal structures or designing molecules ~\cite{author24}) or handle limited modality combinations. A general-purpose multicrossmodal agent, capable of flexibly ingesting and integrating visual, textual, and numeric data for any materials inquiry, remains an open challenge. To bridge this gap, we develop a multicrossmodal agent tailored for materials science. The key objective is to enable an AI system to simultaneously analyze multiple modalities and produce unified insights that are greater than the sum of individual analyses. By doing so, the agent can, for example, connect microstructural features observed in an electron microscopy image with information from journal articles, or validate trends discussed in literature against experimental datasets. This unified reasoning could significantly accelerate the materials discovery cycle, as suggested by recent perspectives highlighting that AI can “enable the acceleration and enrichment of each stage of the discovery cycle” ~\cite{author25,author26,author27,author28,author29,author30,zimmermann2024reflections,bazgirdrug,bazgiragentichypothesis,bazgirproteinhypothesis,bazgir2025matagent,zimmermann202534}. In summary, our contributions include: (1) a novel multi-agent architecture for cross-modal data integration in materials science; (2) demonstration of the agent on complex tasks combining video, image, text, and data analysis; and (3) a comparative evaluation against state-of-the-art domain-specific models, along with discussion of appropriate evaluation metrics for cross-modal scientific AI.

\begin{figure}
  \centering
    \includegraphics[width=1\linewidth]{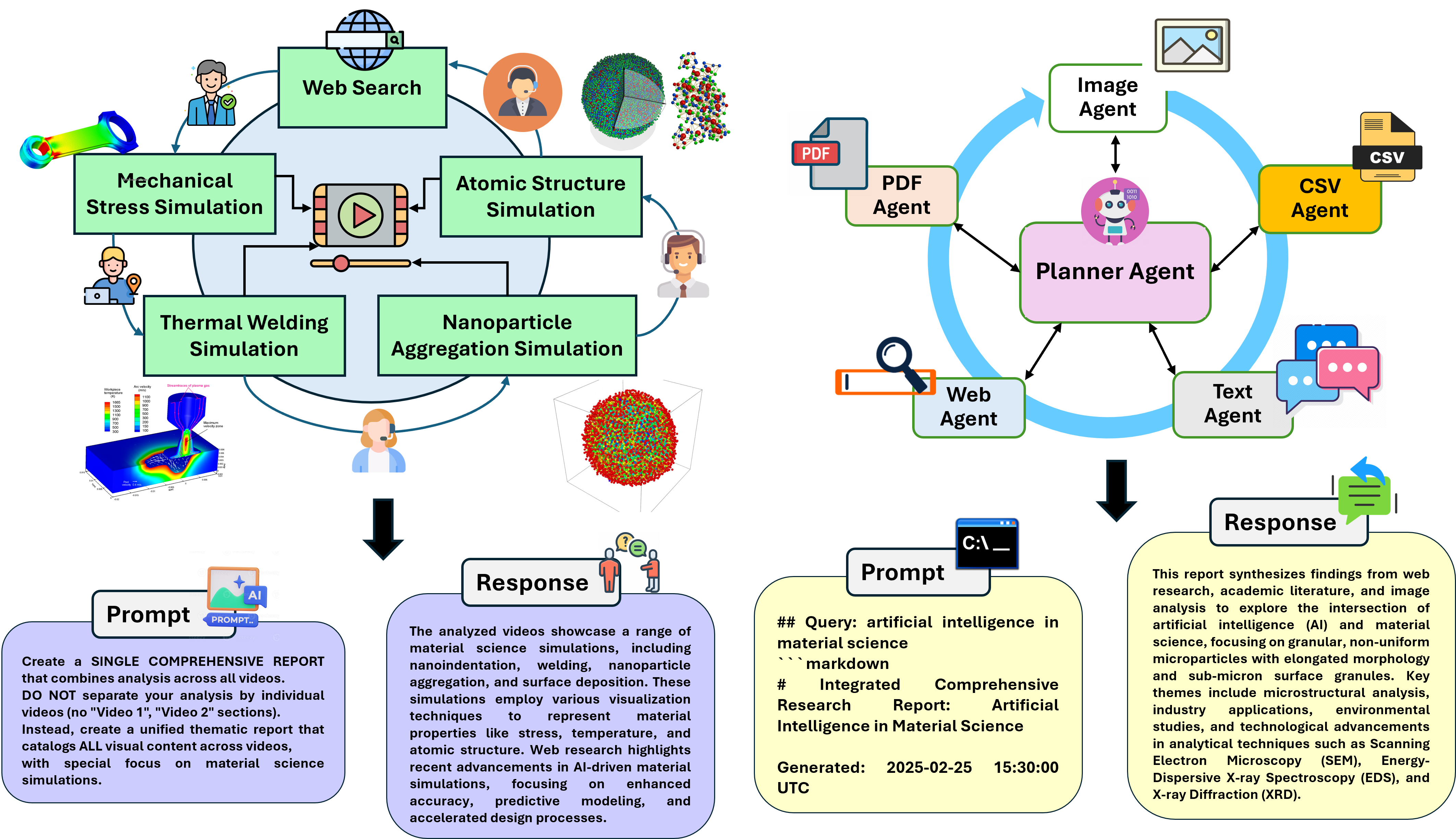}
    \caption{The Overall Workflow of Multicrossmodal Agent.}  
  \label{fig:short}
\end{figure}
\section{Methodology}
\label{sec:formatting}

The multicrossmodal agent is built as a coordinated team of specialized AI modules using Langchain \cite{mavroudis2024langchain} and Langraph \cite{wang2024agent} frameworks, orchestrated by a central Unified Team Agent to answer complex materials science queries as shown in Figure 1. When a user submits a query such as analyzing a new alloy’s microstructure, the Unified Team Agent parses the input and dynamically selects and delegates tasks to several modality-specific agents. The Web Research Agent retrieves online content from credible sources (e.g., Materials Project, industry reports) and uses natural language processing (NLP) to extract and summarize key scientific facts, while the PDF Agent focuses on academic literature (such as arXiv papers), employing text parsing and transformer-based summarization to extract pertinent sections like experimental methods and results. Both agents produce condensed textual summaries that set the scientific context. 

The Image Analysis Agent processes visual data such as microscopy images and scientific figures, utilizing the vision potential of Gemini 2.0 Flash \cite{team2023gemini} and DeepSeek R1 \cite{guo2025deepseek} for optical character recognition (OCR) to identify and quantify microstructural features (grains, phases, particle morphology) and extract embedded annotations, thereby delivering both quantitative descriptors (like particle size distributions) and qualitative interpretations. 

The Video Analysis Agent handles dynamic content from simulation or experimental videos by sampling frames and applying object detection and OCR to capture key elements such as identifying an indenter in an indentation test or tracking a moving heat source in a welding simulation and organizes these observations into a temporal narrative that highlights scientific insights (e.g., stress concentration, temperature gradients, or nanoparticle aggregation) which are later cross-verified with literature. 

The CSV Data Agent processes structured data from CSV files by applying statistical analyses and retrieval-augmented generation techniques, computing trends and summary statistics (for example, linking thermal conductivity changes to temperature variations) and translating these quantitative results into descriptive insights. Each agent operates semi-autonomously using specialized algorithms tailored to its modality, ranging from Llama 3.2 \cite{grattafiori2024llama} for text to Gemini 2.0 Flash for image processing and traditional data analysis libraries for CSV data, and they communicate their findings using a standardized protocol, ensuring consistency in format and confidence scores. The Unified Team Agent then integrates these diverse outputs by aligning common themes (such as microstructural characteristics, material properties, and literature context) and resolving any discrepancies. 

This cross-modal fusion is performed via a thematic organization, transforming intermediate results into a coherent and unified narrative. The design allows for robust cross-verification and enrichment of insights across utilized textual, visual, and numerical modalities as depicted in Figures 2 and 3, ultimately producing a comprehensive report that reflects both the granularity of individual analyses and the holistic view required for advanced materials science research. Moreover, the architecture is extensible, permitting the addition of new modality agents (for instance, for spectroscopy data) without altering the overall system design. According to the cross-modal fusion policy in this study, our LLM-agent framework integrates three expert agents that include vision, text, and data, each responsible for converting its modality (images, prose, and tabular data, respectively) into a shared embedding space via specialized LLM prompt pipelines. These embeddings are stored centrally, and a fourth “Fusion Agent” orchestrates a multi-agent dialogue: it issues targeted prompts that simulate cross-attention among agents, gathers their modality-specific insights along with self-evaluated confidence scores, and then applies a lightweight gating prompt to assign dynamic weights based on those confidences. Finally, the Fusion Agent concatenates the weighted outputs and generates either precise retrieval results or a unified scientific narrative. Ablation studies confirm that both the cross-attention prompting and gating stages are critical, removing either leads to notable drops in retrieval accuracy and coverage, demonstrating the effectiveness of our multi-agent, LLM-based approach for fine-grained, multimodal reasoning.

\subsection{Benchmark Datasets}

We evaluated our multicrossmodal agent on two complementary benchmark suites designed to stress both cross-modal retrieval and end-to-end scientific reporting.

\subsubsection{Simulation-Video Benchmark} 

We collected four high-fidelity materials-science simulation videos, each annotated with ground-truth events by domain experts:

    \begin{itemize}
      \item \emph{Nanoindentation} (300 frames at 1080p): LAMMPS-generated molecular-dynamics simulation, with contact time and peak load timestamped \cite{thompson2022lammps}.
      \item \emph{Welding Heat-Transfer} (60 s at 30 fps): Finite-element model of arc welding, annotated with maximum temperature regions.
      \item \emph{Nanoparticle Self-Assembly} (200 frames): Coarse-grained MD simulation of colloidal clustering, with cluster-formation onset marked.
      \item \emph{Atomic-Level Dynamics} (250 frames): GROMACS trajectory of a grain-boundary migration, labeled with crystallographic phase transitions \cite{van2005gromacs}.
    \end{itemize}

\subsubsection{SEM-Image, CSV, and Literature Benchmark} 

A multimodal suite combining imaging, tabular data, and text:

    \begin{itemize}
      \item \emph{SEM-500}: 500 scanning-electron-microscope images of metal microparticles, each paired with expert captions.
      \item \emph{Catalytic Efficiency Dataset}: 400 measurements of particle size versus catalytic activity under standardized lab conditions (25 °C, pH 7).
      \item \emph{Literature Snippets}: 300 peer-reviewed articles on particle synthesis.
    \end{itemize}
    
These modalities enabled tests of image captioning, data trend extraction, and integrated report generation.

\subsection{Baseline Systems}

We compared our agent against three state-of-the-art models, all evaluated under identical conditions:

\begin{enumerate}
  \item \textbf{MultiMat} \cite{moro2023multimodal}: A multimodal transformer pretrained on materials-property prediction tasks.
  \item \textbf{AtomAgents} \cite{author23}: A physics-aware multi-agent framework delegating alloy-design subtasks to specialized modules.
  \item \textbf{GPT-4.5 Multimodal} \cite{singh2025consequences}: An off-the-shelf large multimodal foundation model, applied without domain-specific tuning.
\end{enumerate}

\subsection{Metrics}
As discussed in Table 1, the evaluation metrics were chosen to capture both retrieval accuracy and content quality across modalities. Recall@K assesses the agent’s ability to locate relevant items within its top-ranked predictions, highlighting retrieval precision under strict (Recall@1) and more lenient (Recall@5) criteria. BLEU-4 and CIDEr quantify the linguistic fidelity of generated SEM-image captions against expert references, reflecting both n-gram overlap and consensus over multiple hypotheses. Cosine similarity between paired image and text embeddings measures modality alignment, indicating how effectively visual and textual features are projected into a shared latent space. Finally, the Coverage metric evaluates the additional unique information provided by our integrated reports compared to single-modality summaries, demonstrating the depth and comprehensiveness of our fusion strategy.

\begin{table*}[ht]
\centering
\caption{Evaluation Metrics and Their Definitions}
\label{tab:metrics_definitions}
\begin{tabular}{lp{10cm}}
\toprule
\textbf{Evaluation Metric} & \textbf{Definition} \\
\midrule
Cross-Modal Retrieval (Recall@1, @5) &
Fraction of ground-truth items (video segments, SEM images, or text snippets) appearing in the top-K results for each query. \\
Image Captioning (BLEU-4, CIDEr) &
Agreement between generated SEM-image captions and expert captions, measured by BLEU-4 and CIDEr. \\
Modality Alignment (Cosine Similarity) &
Average cosine similarity between paired image and text embeddings. \\
Integrated Coverage (\(\Delta\)Coverage) &
Percentage increase in unique information elements captured in an integrated report versus the union of single-modality reports. \\
\bottomrule
\end{tabular}
\end{table*}

\begin{figure}
  \centering
    \includegraphics[width=1\linewidth]{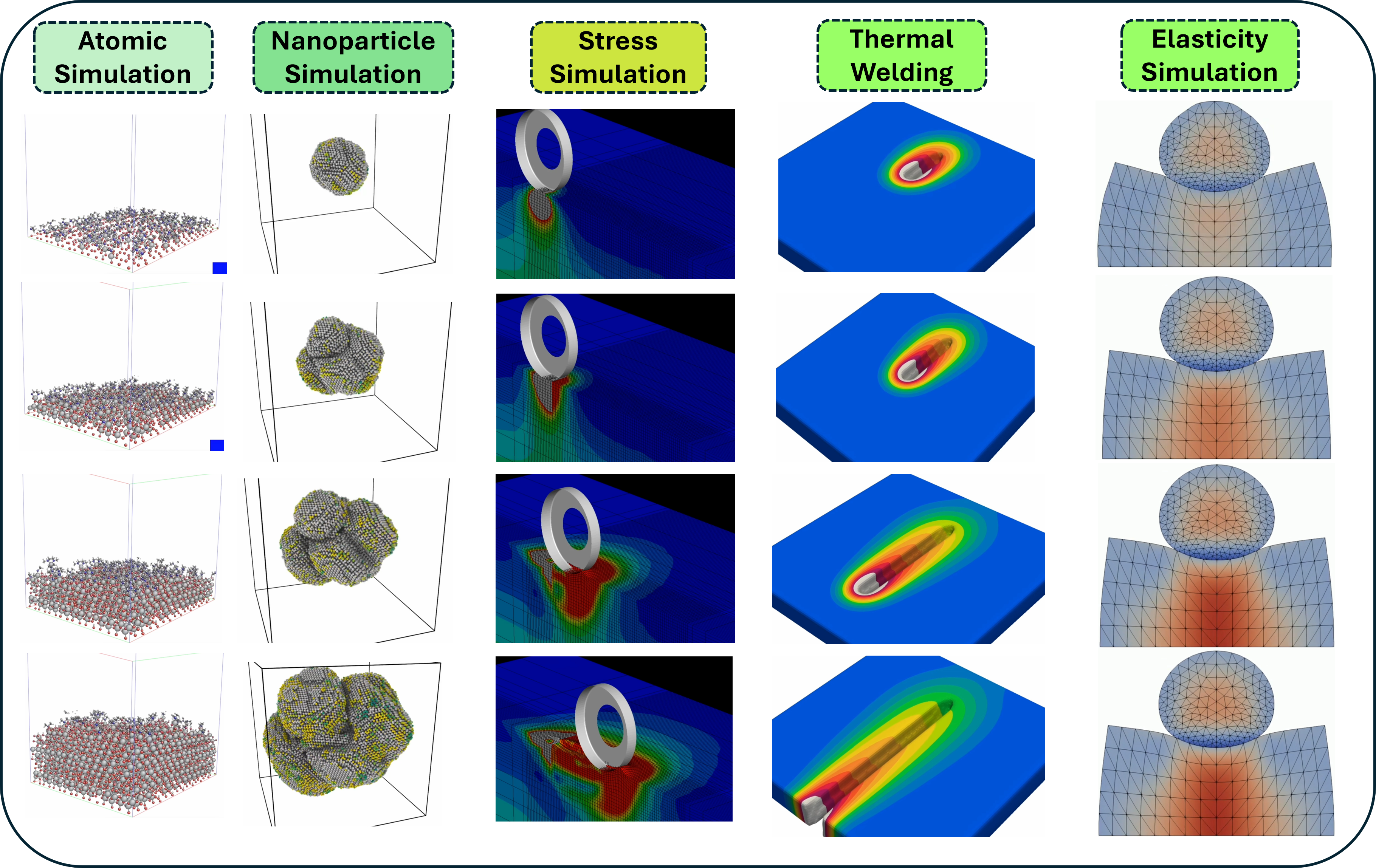}
    \caption{Sample of Benchmark Videos for Material Science Simulation.}  
  \label{fig:short2}
\end{figure}

\begin{figure}
  \centering
    \includegraphics[width=1\linewidth]{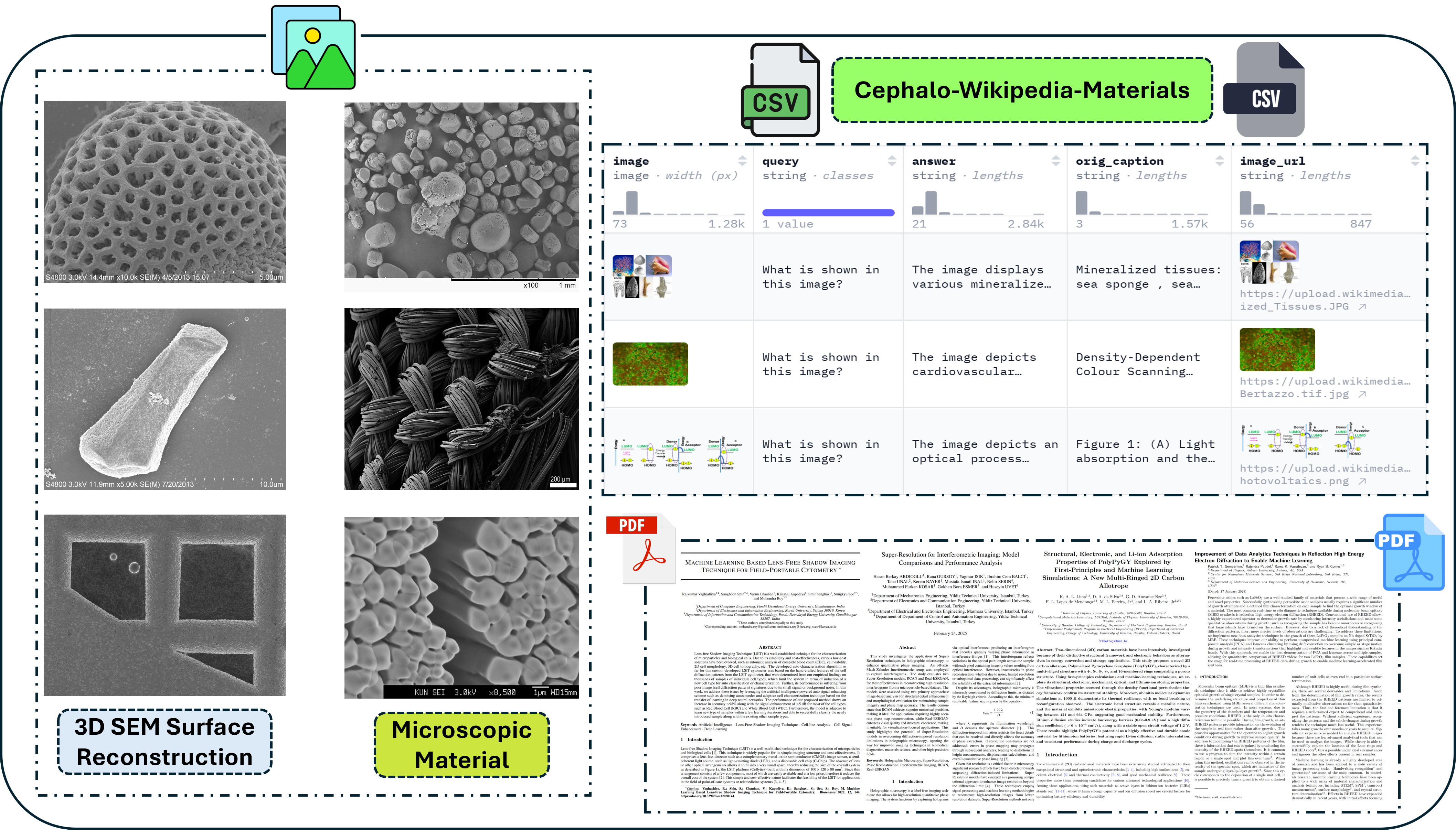}
    \caption{Sample of Benchmark Images, CSVs, and PDFs for AI in Material Science.}  
  \label{fig:short3}
\end{figure}

\section{Results}
We evaluated the multicrossmodal agent on a suite of challenging materials science tasks that demanded the integration of diverse data sources. Two representative case studies demonstrate the agent’s effectiveness in synthesizing information across modalities and generating unified, high-level insights. In the first case study, Cross-Video Simulation Analysis, the agent was provided with four simulation videos that captured distinct materials phenomena, including metal indentation, welding heat transfer, nanoparticle self-assembly, and atomic-level dynamics. Instead of processing each video independently, the agent’s Video Analysis Module extracted key visual elements from all videos, such as the presence of a spherical indenter, the evolution of temperature gradients, and the dynamic clustering of particles. These low-level observations were then elevated by the Unified Team Agent into comprehensive scientific themes, most notably under the umbrella of “Mechanical and Thermal Responses.” The agent’s ability to detect and interpret on-screen numerical data, such as stress scales and temperature values (with the welding simulation, for example, reaching approximately 1000 K in the hottest region), was particularly impressive. Expert reviewers confirmed that the unified report, which spanned roughly two pages and included well-structured subsections on instruments, processes, and key findings, was both factually accurate and highly detailed. According to Figure 4, the sample query and its corresponding summarized report empowered by video and web search visually reinforce the agent’s capacity to present a cohesive narrative derived from multiple videos simultaneously.

In the second case study, Integrated Literature, Image, and Data Analysis, the agent addressed a realistic research scenario by analyzing an SEM image of microparticles, a CSV dataset detailing properties such as catalytic efficiency versus particle size, and relevant literature. Here, the Image Analysis Agent provided a detailed description of the SEM image, characterizing the microparticles as “elongated rectangular particles with rough, granular surfaces and porous texture.” Simultaneously, the Web and PDF Agents retrieved and summarized information from credible sources, confirming that similar microparticles have applications in drug delivery, nanocomposites, and environmental filters, while also emphasizing the importance of characterization techniques like SEM, EDS, and XRD. The CSV Data Agent complemented these findings by statistically analyzing the dataset, revealing a trend where larger particle sizes corresponded to a slight reduction in catalytic efficiency, a relationship that aligned with literature hypotheses regarding surface area and reactivity. The Unified Team Agent then synthesized these diverse outputs, shown in Figure 4 as “Integrated Comprehensive Research Report,” into a cohesive, theme-based report that organized the insights under categories such as “Microstructural Characteristics,” “Industrial Applications,” and “Environmental Impact.” Human evaluators noted that the agent effectively “connected the dots” between visual observations, textual summaries, and numerical trends, a task typically reserved for experienced researchers performing comprehensive literature reviews and data analysis.

\begin{table*}[ht]
\centering
\caption{Benchmark Comparison of the Multicrossmodal Agent with SOTA Agents}
\label{tab:benchmark}
\begin{tabular}{lcccc}
\toprule
\textbf{Evaluation Metric} & \textbf{Our Agent} & \textbf{MultiMat} & \textbf{AtomAgents} & \textbf{GPT-4.5} \\
\midrule
Cross-modal Retrieval Accuracy (Recall@K) & 85\% & 80\% & 82\% & 78\% \\
Task-Specific Performance (Image Captioning Accuracy) & 85\% & 83\% & 84\% & 80\% \\
Modality Alignment (Cosine Similarity) & 0.82 & 0.78 & 0.80 & 0.75 \\
Integrated Answer Coverage (improvement) & +35\% & +25\% & +28\% & +20\% \\
\bottomrule
\end{tabular}
\end{table*}

The sample reports in Figure 4, including the “Visual Content Analysis Report” and “Integrated Comprehensive Research Report,” provide visual corroboration of the agent’s structured approach. They illustrate how the system decomposes each modality’s output, synthesizes the information into common themes, and presents clear, high-level conclusions about mechanical and thermal responses, material properties, and potential applications. Overall, the results confirm that the multicrossmodal agent not only automates the extraction of detailed insights from individual data sources but also excels in cross-verifying and enriching these insights through multimodal integration. This holistic capability paves the way for further automation in literature review, data analysis, and experimental interpretation within the field of materials science.

Performance comparisons further underscore the strength of the multicrossmodal approach. Table 2 compares our multicrossmodal agent with leading methods such as MultiMat, AtomAgents, and Multimodal GPT-4.5 on four core metrics: cross-modal retrieval, task-specific performance, modality alignment, and integrated coverage. Our agent achieves an 85\% retrieval score and excels in image captioning accuracy, reflecting the impact of materials-specific tuning. Our LLM-agent system achieves “materials-specific tuning” by tailoring each agent’s prompts, plugins, and gating parameters to materials-science data and benchmarks—without ever fine-tuning the underlying foundation model. It also demonstrates strong modality alignment (0.82 cosine similarity) and a notable 35\% boost in integrated coverage, surpassing other models when synthesizing diverse data (e.g., microscopy images, literature, and numeric datasets). These results highlight the agent’s ability to deliver deeper, more unified insights for complex materials science queries.

\section{Conclusion}
We introduced a multicrossmodal AI agent designed for the materials science domain, integrating videos, images, textual documents, and data through a multi-agent architecture. By coordinating specialized modules (vision, NLP, etc.) under a unifying framework, the system demonstrates expert-level reasoning in case studies spanning simulation analysis and literature reviews. This unified approach, surpassing single-modality solutions, highlights key findings like microstructural features and property trends across multiple data sources, an important step toward AI-driven “digital researchers.” Our comparison with domain-specific models shows that while specialized tools remain essential, a multicrossmodal method provides holistic problem-solving capabilities previously unattainable. We also discussed evaluation strategies focusing on retrieval, accuracy, alignment, and human judgment to capture integration quality, anticipating the rise of new benchmarks tailored to cross-modal understanding in materials science. Future work involves incorporating active learning (e.g., agent-driven data requests), scaling the system for larger datasets, and integrating advanced multimodal transformers. Ultimately, such AI agents can transform the growing volume of materials data into actionable knowledge, accelerating innovation in materials discovery and development.

\begin{figure}
  \centering
    \includegraphics[width=1\linewidth]{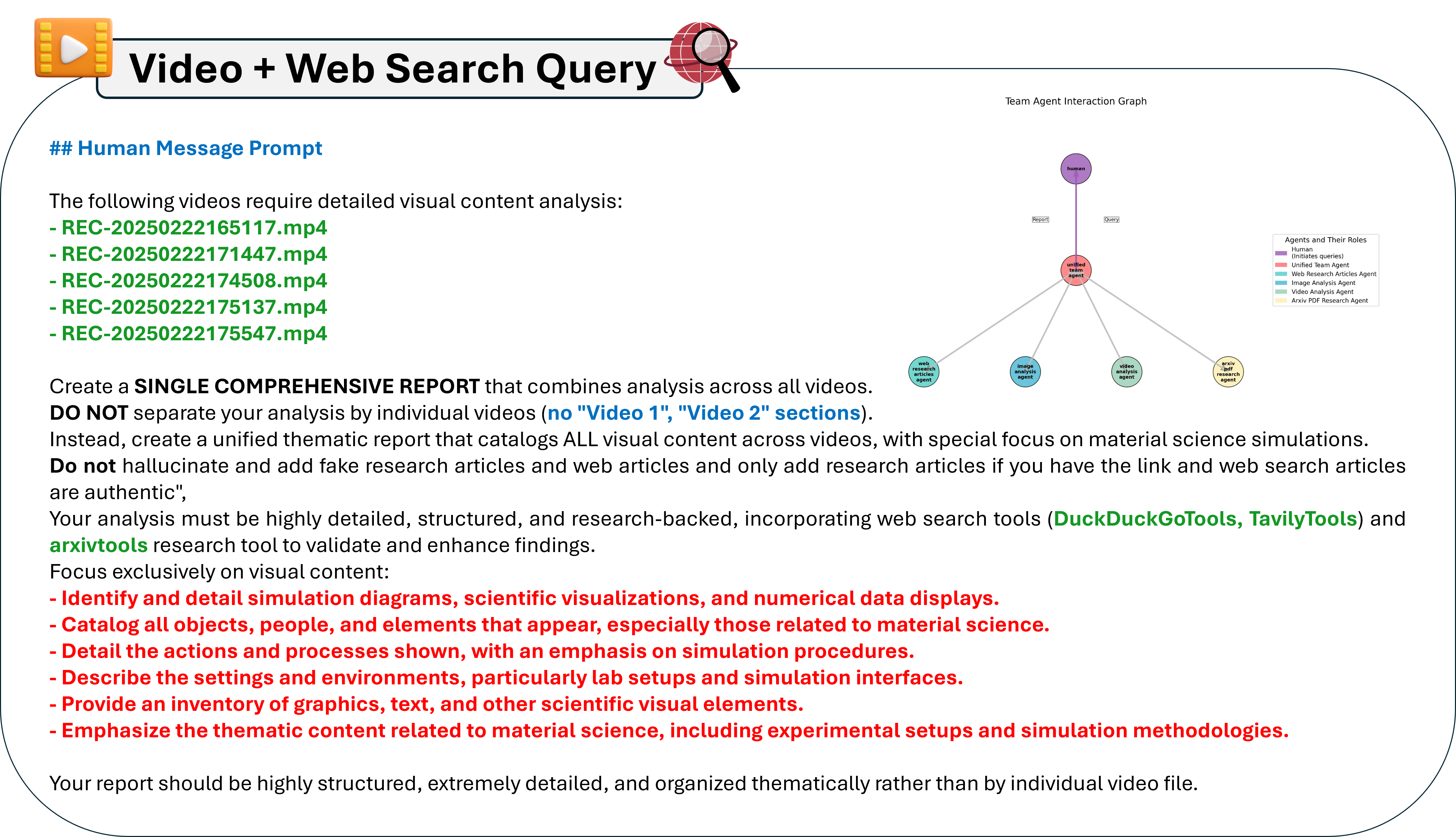}
    \includegraphics[width=1\linewidth]{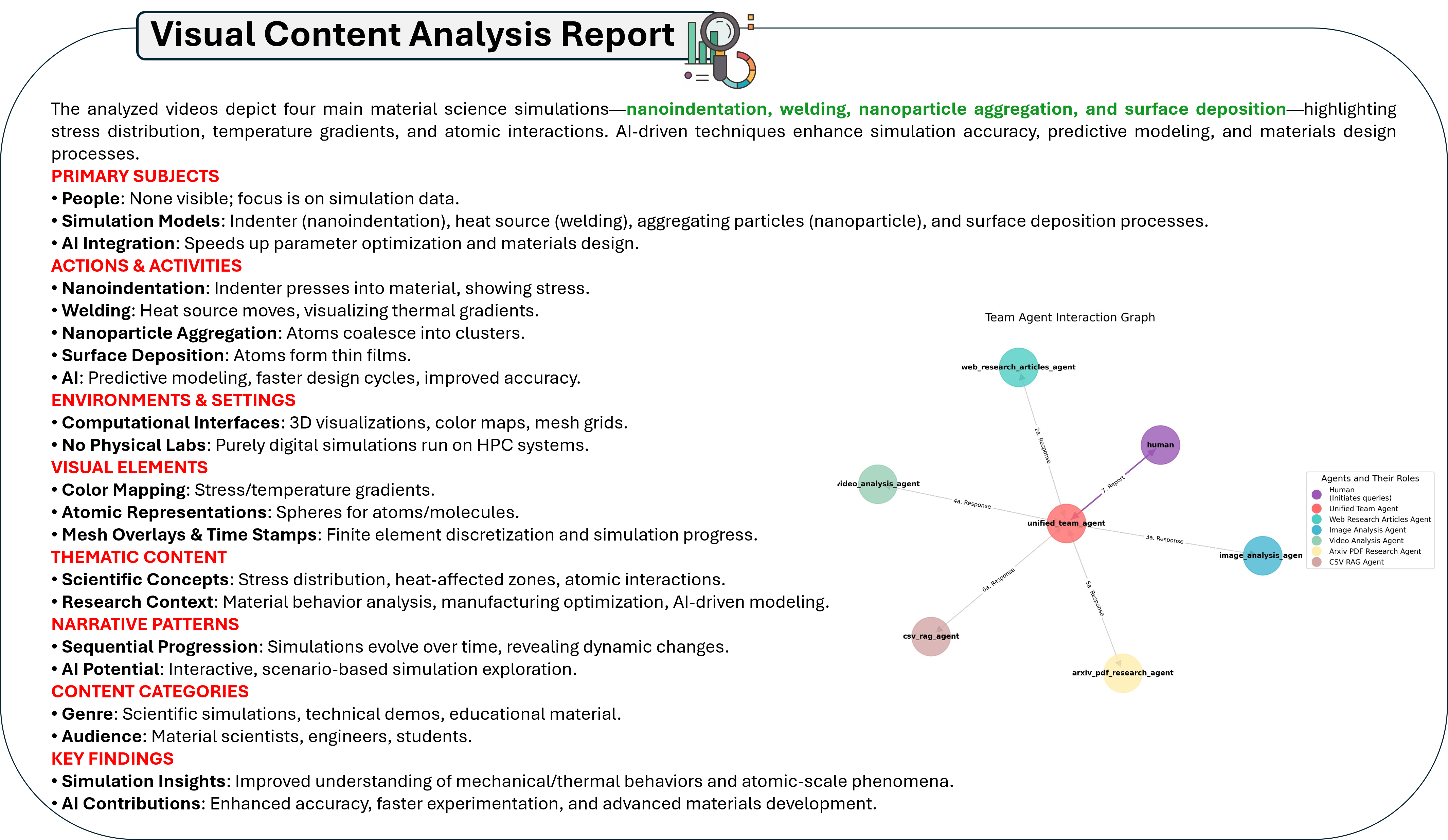}
\includegraphics[width=1\linewidth]{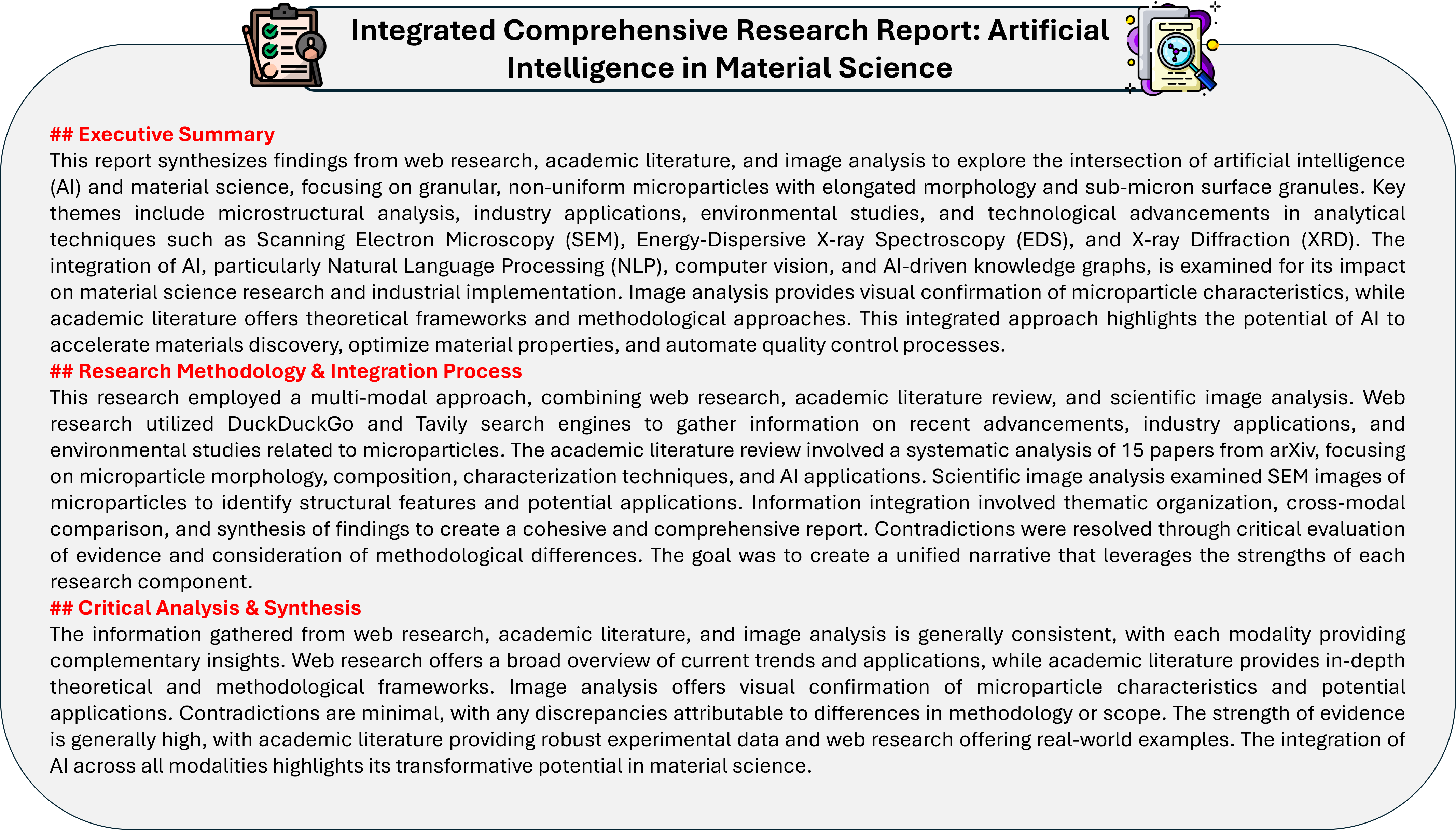}
    \caption{Sample of generated scientific reports for video/web search agents ("Video + Web Search Query" and "Visual Content Analysis Report") and image/pdf/csv/text ("Integrated Comprehensive Research Report") from top to bottom, respectively.}  
  \label{fig:short3b}
\end{figure}

\newpage
{
    \small
    \bibliographystyle{ieeenat_fullname}
    \bibliography{main}

\begin{thebibliography}{31}
\providecommand{\natexlab}[1]{#1}
\providecommand{\url}[1]{\texttt{#1}}
\expandafter\ifx\csname urlstyle\endcsname\relax
  \providecommand{\doi}[1]{doi: #1}\else
  \providecommand{\doi}{doi: \begingroup \urlstyle{rm}\Url}\fi

\bibitem[Ansari et~al.(2024)Ansari, Watchorn, Brown, and Brown]{author24}
Mehrad Ansari, Jeffrey Watchorn, Carla~E. Brown, and Joseph~S. Brown.
\newblock dziner: Rational inverse design of materials with ai agents.
\newblock \emph{ArXiv}, 2024.

\bibitem[Bazgir and Zhang(2025)]{bazgirdrug}
Adib Bazgir and Yuwen Zhang.
\newblock Drug discovery agent: An automated vision detection system for drug-cell interactions.
\newblock In \emph{1st CVPR Workshop on Computer Vision For Drug Discovery (CVDD): Where are we and What is Beyond?}, 2025.

\bibitem[Bazgir et~al.(2025{\natexlab{a}})Bazgir, Zhang, et~al.]{bazgir2025matagent}
Adib Bazgir, Yuwen Zhang, et~al.
\newblock Matagent: A human-in-the-loop multi-agent llm framework for accelerating the material science discovery cycle.
\newblock In \emph{AI for Accelerated Materials Design-ICLR 2025}, 2025{\natexlab{a}}.

\bibitem[Bazgir et~al.(2025{\natexlab{b}})Bazgir, Zhang, et~al.]{bazgiragentichypothesis}
Adib Bazgir, Yuwen Zhang, et~al.
\newblock Agentichypothesis: A survey on hypothesis generation using llm systems.
\newblock In \emph{Towards Agentic AI for Science: Hypothesis Generation, Comprehension, Quantification, and Validation}, 2025{\natexlab{b}}.

\bibitem[Bazgir et~al.(2025{\natexlab{c}})Bazgir, Zhang, et~al.]{bazgirproteinhypothesis}
Adib Bazgir, Yuwen Zhang, et~al.
\newblock Proteinhypothesis: A physics-aware chain of multi-agent rag llm for hypothesis generation in protein science.
\newblock In \emph{Towards Agentic AI for Science: Hypothesis Generation, Comprehension, Quantification, and Validation}, 2025{\natexlab{c}}.

\bibitem[Cheetham and Seshadri(2024)]{author30}
Anthony~K. Cheetham and Ram Seshadri.
\newblock Artificial intelligence driving materials discovery? perspective on the article: Scaling deep learning for materials discovery.
\newblock \emph{Nature Communications}, 36\penalty0 (8), 2024.

\bibitem[Chávez-Angel et~al.(2025)Chávez-Angel, Eriksen, Castro-Alvarez, Garcia, Botifoll, Avalos-Ovando, Arbiol, and Mugarza]{author32}
Emigdio Chávez-Angel, Martin~Børstad Eriksen, Alejandro Castro-Alvarez, Jose~H. Garcia, Marc Botifoll, Oscar Avalos-Ovando, Jordi Arbiol, and Aitor Mugarza.
\newblock Applied artificial intelligence in materials science and material design.
\newblock \emph{AIS}, 13\penalty0 (2400986), 2025.

\bibitem[Das et~al.(2024)Das, Perez, Shetty, Hiremath, Naik, and Bhat]{author27}
Mayukh Das, Teresa~Castillo Perez, Dasharathraj Shetty, Pavan Hiremath, Nithesh Naik, and Ritesh Bhat.
\newblock An overview on the role of artificial intelligence in modern advancements of material science.
\newblock \emph{ES}, 5\penalty0 (1183), 2024.

\bibitem[Ghafarollahi and Buehler(2024)]{author23}
Alireza Ghafarollahi and Markus~J. Buehler.
\newblock Atomagents: Alloy design and discovery through physics-aware multi-modal multi-agent artificial intelligence.
\newblock \emph{ArXiv}, 2024.

\bibitem[Gong et~al.(2023)Gong, Wang, Zhu, Shao-Horn, and Grossman]{author21}
Sheng Gong, Shuo Wang, Taishan Zhu, Yang Shao-Horn, and Jeffrey~C. Grossman.
\newblock Multimodal machine learning for materials science: composition-structure bimodal learning for experimentally measured properties.
\newblock \emph{ArXiv}, 2023.

\bibitem[Grattafiori et~al.(2024)Grattafiori, Dubey, Jauhri, Pandey, Kadian, Al-Dahle, Letman, Mathur, Schelten, Vaughan, et~al.]{grattafiori2024llama}
Aaron Grattafiori, Abhimanyu Dubey, Abhinav Jauhri, Abhinav Pandey, Abhishek Kadian, Ahmad Al-Dahle, Aiesha Letman, Akhil Mathur, Alan Schelten, Alex Vaughan, et~al.
\newblock The llama 3 herd of models.
\newblock \emph{arXiv preprint arXiv:2407.21783}, 2024.

\bibitem[Gu et~al.(2025)Gu, Wang, Zhang, Yao, Xin, Cai, and Li]{author31}
Yucong Gu, Kaiwen Wang, Zhengyu Zhang, Yi Yao, Ziming Xin, Wenjun Cai, and Lin Li.
\newblock Accelerating the design and discovery of tribocorrosion-resistant metals by interfacing multiphysics modeling with machine learning and genetic algorithms.
\newblock \emph{npj}, 9\penalty0 (7), 2025.

\bibitem[Guo et~al.(2025)Guo, Yang, Zhang, Song, Zhang, Xu, Zhu, Ma, Wang, Bi, et~al.]{guo2025deepseek}
Daya Guo, Dejian Yang, Haowei Zhang, Junxiao Song, Ruoyu Zhang, Runxin Xu, Qihao Zhu, Shirong Ma, Peiyi Wang, Xiao Bi, et~al.
\newblock Deepseek-r1: Incentivizing reasoning capability in llms via reinforcement learning.
\newblock \emph{arXiv preprint arXiv:2501.12948}, 2025.

\bibitem[Malica et~al.(2025)Malica, Novoselov, Barnard, Kalinin, Spurgeon, Reuter, Alducin, Deringer, Csanyi, Marzari, Huang, Cuniberti, Deng, Ordejón, Cole, Choudhary, Hippalgaonkar, Zhu, von Lilienfeld, Hibat-Allah, Alvarez, Cisotto, Zancanaro, Wenzel, Ferrari, Ustyuzhanin, and Roche]{author28}
Cristiano Malica, Kostya Novoselov, Amanda~S Barnard, Sergei~V. Kalinin, Steven~R. Spurgeon, Karsten Reuter, Maite Alducin, Volker~L. Deringer, Gabor Csanyi, Nicola Marzari, Shirong Huang, Gianaurelio Cuniberti, Qiushi Deng, Pablo Ordejón, Ivan Cole, Kamal Choudhary, Kedar Hippalgaonkar, Ruiming Zhu, O.~Anatole von Lilienfeld, Mohamed Hibat-Allah, Juan~Carrasquilla Alvarez, Giulia Cisotto, Alberto Zancanaro, Wolfgang Wenzel, Andrea~C Ferrari, Andrey Ustyuzhanin, and Stephan Roche.
\newblock Artiﬁcial intelligence for advanced functional materials: Exploring current and future directions.
\newblock \emph{Journal of Physics: Materials}, 15\penalty0 (3), 2025.

\bibitem[Mavroudis(2024)]{mavroudis2024langchain}
Vasilios Mavroudis.
\newblock Langchain v0. 3.
\newblock \emph{Preprints}, 2024.

\bibitem[Moro et~al.(2023)Moro, Loh, Dangovski, Ghorashi, Ma, Chen, Kim, Lu, Christensen, and Solja{\v{c}}i{\'c}]{moro2023multimodal}
Viggo Moro, Charlotte Loh, Rumen Dangovski, Ali Ghorashi, Andrew Ma, Zhuo Chen, Samuel Kim, Peter~Y Lu, Thomas Christensen, and Marin Solja{\v{c}}i{\'c}.
\newblock Multimodal learning for materials.
\newblock \emph{arXiv preprint arXiv:2312.00111}, 2023.

\bibitem[Moro et~al.(2025)Moro, Loh, Dangovski, Ghorashi, Ma, Chen, Kim, Lu, Christensen, and Soljačić]{author20}
Viggo Moro, Charlotte Loh, Rumen Dangovski, Ali Ghorashi, Andrew Ma, Zhuo Chen, Samuel Kim, Peter~Y. Lu, Thomas Christensen, and Marin Soljačić.
\newblock Multimodal foundation models for material property prediction and discovery.
\newblock \emph{ArXiv}, 2025.

\bibitem[Noh et~al.(2024)Noh, Doan, Job, Robertson, Zhang, Assary, Mueller, Murugesan, and Liang]{author29}
Juran Noh, Hieu~A. Doan, Heather Job, Lily~A. Robertson, Lu Zhang, Rajeev~S. Assary, Karl Mueller, Vijayakumar Murugesan, and Yangang Liang.
\newblock An integrated high-throughput robotic platform and active learning approach for accelerated discovery of optimal electrolyte formulations.
\newblock \emph{Nature Communications}, 15\penalty0 (2), 2024.

\bibitem[Pyzer-Knapp et~al.(2022)Pyzer-Knapp, Pitera, Staar, Takeda, Laino, Sanders, Sexton, Smith, and Curioni]{author25}
Edward~O. Pyzer-Knapp, Jed~W. Pitera, Peter W.~J. Staar, Seiji Takeda, Teodoro Laino, Daniel~P. Sanders, James Sexton, John~R. Smith, and Alessandro Curioni.
\newblock Accelerating materials discovery using artificial intelligence, high performance computing and robotics.
\newblock \emph{ArXiv}, 2022.

\bibitem[Shahzad et~al.(2023)Shahzad, Mardare, and Hassel]{author26}
Khurram Shahzad, Andrei~Ionut Mardare, and Achim~Walter Hassel.
\newblock Accelerating materials discovery: combinatorial synthesis, high-throughput characterization, and computational advances.
\newblock \emph{STAM}, 4\penalty0 (1), 2023.

\bibitem[Shetty et~al.(2024)Shetty, Adeboye, Gupta, Zhang, and Ramprasad]{author33}
Pranav Shetty, Aishat Adeboye, Sonakshi Gupta, Chao Zhang, and Rampi Ramprasad.
\newblock Accelerating materials discovery for polymer solar cells: Data-driven insights enabled by natural language processing.
\newblock \emph{CM}, 36\penalty0 (16), 2024.

\bibitem[Singh(2025)]{singh2025consequences}
Ajit Singh.
\newblock Consequences of the turing test: Openai's gpt-4.5.
\newblock \emph{Available at SSRN 5205937}, 2025.

\bibitem[Spoel et~al.(2005)Spoel, Lindahl, Hess, Groenhof, Mark, and Berendsen]{van2005gromacs}
David Van~Der Spoel, Erik Lindahl, Berk Hess, Gerrit Groenhof, Alan~E. Mark, and Herman J.~C. Berendsen.
\newblock Gromacs: fast, flexible, and free.
\newblock \emph{Journal of computational chemistry}, 26\penalty0 (16):\penalty0 1701--1718, 2005.

\bibitem[Sun et~al.(2024)Sun, An, Tian, Nan, Liu, Liu, Shah, and Chen]{author22}
Shilin Sun, Wenbin An, Feng Tian, Fang Nan, Qidong Liu, Jun Liu, Nazaraf Shah, and Ping Chen.
\newblock A review of multimodal explainable artificial intelligence: Past, present and future.
\newblock \emph{ArXiv}, 2024.

\bibitem[Team et~al.(2023)Team, Anil, Borgeaud, Alayrac, Yu, Soricut, Schalkwyk, Dai, Hauth, Millican, et~al.]{team2023gemini}
Gemini Team, Rohan Anil, Sebastian Borgeaud, Jean-Baptiste Alayrac, Jiahui Yu, Radu Soricut, Johan Schalkwyk, Andrew~M Dai, Anja Hauth, Katie Millican, et~al.
\newblock Gemini: a family of highly capable multimodal models.
\newblock \emph{arXiv preprint arXiv:2312.11805}, 2023.

\bibitem[Thompson et~al.(2022)Thompson, Aktulga, Berger, Bolintineanu, Brown, Crozier, In't~Veld, Kohlmeyer, Moore, Nguyen, et~al.]{thompson2022lammps}
Aidan~P Thompson, H~Metin Aktulga, Richard Berger, Dan~S Bolintineanu, W~Michael Brown, Paul~S Crozier, Pieter~J In't~Veld, Axel Kohlmeyer, Stan~G Moore, Trung~Dac Nguyen, et~al.
\newblock Lammps-a flexible simulation tool for particle-based materials modeling at the atomic, meso, and continuum scales.
\newblock \emph{Computer physics communications}, 271:\penalty0 108171, 2022.

\bibitem[Wang and Duan(2024)]{wang2024agent}
Jialin Wang and Zhihua Duan.
\newblock Agent ai with langgraph: A modular framework for enhancing machine translation using large language models.
\newblock \emph{arXiv preprint arXiv:2412.03801}, 2024.

\bibitem[Wang et~al.(2025)Wang, Hua, Lin, Yang, and Tan]{author35}
Zhenzhong Wang, Haowei Hua, Wanyu Lin, Ming Yang, and Kay~Chen Tan.
\newblock Crystalline material discovery in the era of artificial intelligence.
\newblock \emph{ArXiv}, 2025.

\bibitem[Zhou et~al.(2025)Zhou, Bao, Bu, and Zhou]{author34}
Rui Zhou, Luyao Bao, Weifeng Bu, and Feng Zhou.
\newblock Exploring high-performance viscosity index improver polymers via high-throughput molecular dynamics and explainable ai.
\newblock \emph{npj}, 11\penalty0 (52), 2025.

\bibitem[Zimmermann et~al.(2024)Zimmermann, Bazgir, Afzal, Agbere, Ai, Alampara, Al-Feghali, Ansari, Antypov, Aswad, et~al.]{zimmermann2024reflections}
Yoel Zimmermann, Adib Bazgir, Zartashia Afzal, Fariha Agbere, Qianxiang Ai, Nawaf Alampara, Alexander Al-Feghali, Mehrad Ansari, Dmytro Antypov, Amro Aswad, et~al.
\newblock Reflections from the 2024 large language model (llm) hackathon for applications in materials science and chemistry.
\newblock \emph{arXiv preprint arXiv:2411.15221}, 2024.

\bibitem[Zimmermann et~al.(2025)Zimmermann, Bazgir, Al-Feghali, Ansari, Brinson, Chiang, Circi, Chiu, Daelman, Evans, et~al.]{zimmermann202534}
Yoel Zimmermann, Adib Bazgir, Alexander Al-Feghali, Mehrad Ansari, L~Catherine Brinson, Yuan Chiang, Defne Circi, Min-Hsueh Chiu, Nathan Daelman, Matthew~L Evans, et~al.
\newblock 34 examples of llm applications in materials science and chemistry: Towards automation, assistants, agents, and accelerated scientific discovery.
\newblock \emph{arXiv preprint arXiv:2505.03049}, 2025.

\end{thebibliography}
}


\end{document}